\documentclass[prb,floatfix,superscriptaddress,twocolumn,aps,showpacs,preprintnumbers,amsmath,amssymb]{revtex4}
\usepackage[latin1]{inputenc}
\usepackage{graphicx}
\usepackage{psfrag}
\usepackage{wasysym}
\graphicspath{/net/inttheorie/home/fevers/Tex/Pap.Mat/p3.surfMelting/werner/}

\usepackage{ulem}

\newcommand{\br}{{\boldsymbol r}}

\begin{document}

\title{Lowering of surface melting temperature in atomic clusters \\ with a
  nearly closed shell structure}
\author{ A. Bagrets}    
  \affiliation{Institut f\"ur Nanotechnologie, Forschungszentrum Karlsruhe, D-76021 Karlsruhe,
    Germany}
\author{ R. Werner\footnote{Present address: d-fine GmbH, Opernplatz 2, 60313 Frankfurt,
  Germany}
}    
\affiliation{\mbox{Institut f\"ur Theorie der Kondensierten Materie,
  Universit\"at Karlsruhe, D-76128 Karlsruhe, Germany}}
\author{F. Evers}%
  \affiliation{\mbox{Institut f\"ur Theorie der Kondensierten Materie, Universit\"at Karlsruhe,
    D-76128 Karlsruhe, Germany}} 
  \affiliation{Institut f\"ur Nanotechnologie, Forschungszentrum Karlsruhe, D-76021 Karlsruhe,
    Germany}
\author{G. Schneider\footnote{Present address: Department of Physics, Oregon State University,
    301 Weniger Hall Corvallis, OR 97331-6507, USA}
}%
  \affiliation{\mbox{Institut f\"ur Theorie der Kondensierten Materie, Universit\"at Karlsruhe,
    D-76128 Karlsruhe, Germany}} 
\author{D. Schooss}%
  \affiliation{Institut f\"ur Nanotechnologie, Forschungszentrum Karlsruhe, D-76021 Karlsruhe,
    Germany}
\author{P. W\"olfle}%
  \affiliation{\mbox{Institut f\"ur Theorie der Kondensierten Materie, Universit\"at Karlsruhe,
    D-76128 Karlsruhe, Germany}} 
  \affiliation{Institut f\"ur Nanotechnologie, Forschungszentrum Karlsruhe, D-76021 Karlsruhe,
    Germany}

\date{\today}

\begin{abstract}
We investigate the interplay of particle number, $N$, and structural
properties of selected clusters
with $N{=}12$ up to $N{=}562$ 
by employing Gupta potentials parameterized for Aluminum
and extensive Monte-Carlo simulations. 
Our analysis focuses on closed shell structures
with extra atoms. The latter 
can put the cluster under a significant stress and we argue that typically such a 
strained system exhibits a reduced energy barrier for 
(surface) diffusion of cluster atoms. Consequently, also
its surface melting temperature,
$T_{\rm S}$, is reduced, 
so that $T_{\rm S}$ separates from and actually falls 
well below the bulk value. The proposed mechanism may be responsible 
for the suppression of the surface melting temperature observed in 
a recent experiments. 
\end{abstract}
 
\maketitle

\section{Introduction}

The properties of small metal clusters have enjoyed considerable interest
in recent years. Cluster properties can differ enormously from 
those of the bulk material due to the  large surface-to-volume ratio and
due to a remarkable sensitivity of the electronic structure to the
cluster size and geometry.\cite{hakkinen03} 
These properties are of interest in 
technological applications, e.~g.\ for catalysis.
\cite{catalysis03} From a
conceptual point of view clusters pose 
fundamental questions in statistical
mechanics of finite systems.\cite{gross01}

The melting process of small clusters is a 
complex phenomenon, the detailed rules of which are poorly 
understood. From early on it has been 
associated with isomer fluctuations \cite{AB86,JBB86}. More recent
investigations of {\it isolated} Ni$_{13-x}$Al$_x$ alloy clusters 
\cite{KJ97} elucidate the relation between isomer
fluctuations and the increase of entropy across the melting
transition. 
A detailed overview about how the phase space grows with
increasing particle number and the classification of isomers in terms
of potential energy surfaces is given in Ref.~\onlinecite{WDM+00}. A
general overview of structural properties of nanoclusters is found in
Ref.~\onlinecite{BF05}.

By contrast, the understanding of the melting of very large 
clusters and bulk materials is considerably better developed. 
It is believed to be a strongly inhomogeneous process. 
In large, free metal clusters melting starts in the outermost atom shell, 
at the interface to the vacuum, 
because there the thermal fluctuations of the particle density 
have the lowest energy cost.\cite{GS70,cahn86}  
By feeding more heat into the cluster 
melting peals off shells from the solid cluster core, layer by layer. 
Following this logic, one concludes
that melting in principle is a continuous transition -- to the extent
that each one of the individual layers carries its own melting 
temperature. For atomic clusters 
this is the rule rather than the exception,
because typically each layer has its own atomistic structure. 
This behavior is reflected in the temperature dependence of the
specific heat, $C(T)$, which is not necessarily a
very sharply peaked function of temperature
(as is the case for bulk samples), but rather may exhibit a 
strong  inhomogeneous broadening which reflects inter-shell averaging. 
A historical overview on ``continuous melting'' 
is given in Ref.~\onlinecite{Dash02}.

Depending on the crystal orientation \cite{PGFV87}, 
the surface may melt already at temperatures
well {\it below} the bulk melting point \cite{FV85}. The thickness 
of the molten layer is strongly temperature dependent. It increases
continuously with increasing temperature and eventually it diverges --
by definition -- at the bulk melting temperature. 
(It is assumed here, that the thermodynamic state of the
cluster interior is unique, i.~e. it becomes
independent of geometry in the thermodynamic limit.)
These observations can be treated
theoretically in model calculations using
different effective potentials \cite{OLW94,TET95} as
well as phenomenologically \cite{TET95}.

Coming back to small systems, 
this dependency of melting on surface crystallography 
suggests that surface melting phenomena in atomic clusters 
should exhibit pronounced size effects, i.~e. the melting behavior of
two clusters, that differ in size only by one atom, can vary
significantly. In particular, the (atomic) structure of closed (atom) shell 
clusters (magic clusters) is very sensitive 
to the addition of ad-atoms or vacancies
\cite{DW98,AE99,TJW00,aguado06,noya07}. 

Our paper offers a systematic study of selected Al clusters 
in a range $12\leq N\leq 562$ near their melting transition. 
Here, $N$ is small enough, 
so that a simple extrapolation based on a the continuum 
theory is not applicable and new physics should emerge. 
Our most crucial observation formulated in general terms:
consider splitting the free energy of an $N$-atom 
cluster into a bulk and a surface contribution
\begin{equation}
\label{e0}
F(T,N) = N f_{\rm B}(T) + N^{2/3}f_{\rm S}(T). 
\end{equation}
Both terms, $f_{\rm B,S}$, depend on geometrical details of the
cluster, i.~e. we expect them to become strictly independent of $N$
only in the limit $N{\to}\infty$. 
Formally, $f_{\rm B,S}$ are related to two reservoirs, 
called surface and bulk, with their own specific free energies. 
The reservoirs are coupled in the sense of the grand canonical
ensemble, so they can exchange energy and particles. 
From this point of view, there is no reason why surface and bulk
should have the same, or even a similar, melting temperature. 
\cite{KB93}
A reason why in metal clusters both temperatures tend to be 
strongly correlated with one another, nevertheless, 
is that the interatomic forces at 
the surface and inside the bulk are similar. 

Based on Monte-Carlo simulations employing semi-empirical
Gupta potentials, we propose a general
mechanism that can lead to a considerable splitting of the
surface and bulk melting temperatures. 
Consider a closed shell cluster, e.~g. 
Al$_{13}$ or Al$_{55}$, with icosahedral symmetry. 
The outer shell of the Ih$_{55}$ can accommodate an
additional atom, an ``impurity interstitial'', 
by replacing the fivefold ring structure 
surrounding an edge atom by a sixfold {\it rosettelike} ring. 
The formation of rosettelike structural excitations 
has been introduced already as a route to amorphisation of
Ih$_{55}$ systems\cite{apra04}, 
and we propose that is relevant for
the binding of adatoms in Ih$_{56}$ and as well. 

The impurity is mobile at the surface and its 
motion strongly assists surface melting. 
This is, because the atoms inside the meandering deformation field 
are pushed away from their favorite high symmetry, 
low energy site into a more shallow potential well at intermediate position. 
Our explicit calculations strongly suggest, that by this mechanism the
self diffusion of surface atoms can be dramatically enhanced
indicating a significant reduction in the activation barrier for
diffusion and 
similarly also of the surface melting temperature. 
Since impurities do not enter the next (second one counted from
outside to inside) cluster shell, 
there is no corresponding reduction there, so that only two
different melting transitions should be discriminated.  

This effect may have been seen in two recent experiments. 
Haberland {\it et al.}\cite{haberland05,schmidt98} have determined 
the latent heat and the melting  entropy of 
sodium clusters, Na$_{N}$, with $N{\approx}50{-}360$. 
Their modeling of the data provides an excellent phenomenological
description {\it assuming} the premelting of the cluster surface  
for non-magical $N$-values. The microscopic mechanism, that is
responsible for this lowering of the surfaces melting temperature,
remained unspecified, however. In subsequent theoretical work 
the experimental melting temperatures have been reproduced
quantitatively for a selected set of clusters \cite{aguado05a}. 
Moreover, the microscopics of premelting of 
sub-magic clusters has been already understood in terms 
of the diffusion of vacancies \cite{aguado05c,krishnamurty07}. 
Still, the effect of ad-atoms has not been explicitly analyzed, 
and the relation to elasticity theory (strain) remained unexplored.

Also, the specific heat of Aluminum cluster cations
has been measured recently in the interval $49\leq N\leq 63$ 
by multicollision induced dissociation \cite{BNCJ05}. Interestingly, 
the specific heat data for a number of
clusters shows signatures of multiple transitions, which have
tentatively been interpreted as solid-liquid 
transitions at the surface that occur below the onset of
melting.\cite{neal07} 



\section{Method}
To calculate the thermodynamics of an $N$-atom metal cluster, 
we employ a MC simulation in the canonical ensemble. Technical aspects
of our procedure are described in detail in Ref.~\onlinecite{Wern05a}; 
here, we focus on basic conceptual issues in order to provide 
the prerequisites necessary for a careful 
discussion of the numerical observables in Secs.~\ref{sIII} and \ref{sIV}. 

The potential energy of the 
metal cluster can efficiently be modeled by effective 
many-body potentials. We shall employ the Gupta potential (GP) \cite{Gupt81},
which can be derived in the second moment approximation from a tight
binding model \cite{TMB83,ZLT91} and which correctly describes the
surface contraction observed in metals:
\begin{equation}\label{Gupta}
V(\{r_{ij}\}) = \sum_{i}^N\left[\sum_{j\neq i}^N
A e^{-p\, \overline{r}_{ij}} - 
\sqrt{\sum_{j\neq i} \xi^2 e^{-2q\, \overline{r}_{ij}}}\right]\,.
\end{equation}
Here, $i$ and $j$ are atom labels,
$\overline{r}_{ij} = r_{ij}/r_0 - 1$, and $r_{ij} = |\br_i - \br_j|$
is the  modulus of the distance between two atoms at positions
$\br_i$ and $\br_j$. The parameters have been determined by fitting
the experimental bulk lattice parameters and elastic moduli
\cite{CR93} as $A = 0.1221$ eV, $\xi = 1.316$ eV, $p = 8.612$, and $q
= 2.516$ for Al. Distances are measured in units of the bulk first
neighbor distance $r_0 = 2.864$ \AA. 
A standard Metropolis algorithm
is employed \cite{AT89,WDS+01,Wern05a} with boundary conditions
imposed by a hard wall cube with linear dimension $L$:
a shift of a single atom by a randomly chosen vector with a length 
taken from the interval $[0,\kappa(T) r_0]$ is offered with an 
associated change of the cluster energy $\Delta E$.  
A temperature, $T$, 
is introduced via the probability, $p$, to accept such a step 
with $p {\sim}\exp(-\Delta E/k_{\rm B}T)$. The parameter 
$0{<}\kappa(T){<}1$ is chosen
so that the acceptance rate is close to 50\%; a typical value at
intermediate temperatures is $\kappa{=}0.25$. 
The cluster is updated after each
accepted move. Runs are performed with sampling rates 
of up to $8\times10^7$ steps per temperature and atom.

\subsection*{Observables}

In order to characterize the thermodynamic state of the 
cluster we introduce the following observables: 

(i) The specific heat can be obtained from the ensemble averages of
the potential energy $V$ and its square,
\begin{equation}\label{C}
\frac{C}{k_{\rm B}} = \frac{1}{Nk_{\rm B}^2T^2}
    \left(\langle V^2 \rangle - \langle V \rangle^2\right) 
    + \frac{3}{2}\,.
\end{equation}
The kinetic contribution $C_{\rm kin} = 3/2 k_{\rm B}$ per atom has been
added here.
 (Note, that we treat the metal cluster as a gas of
distinguishable particles rather than indistinguishable ones. 
For the specific heat, a double counting problem does not arise, 
since it is a {\it second} derivative of the free energy.)    

Invoking ergodicity, 
in practical calculations the ensemble average
$\langle \ldots \rangle$ is frequently combined with 
(or even replaced by) an average over the 
MC ``time'', $\tau$, i.e. the total number of MC steps: 
\begin{equation}
\label{e5}
\{ {\cal O}_{ij}\}(\tau) = \frac{1}{\tau} \sum_{n{=}1} {\cal
  O}_{ij}^{(n)}, 
\end{equation}
counting is started after equilibrating an initial configuration. 
The two body term ${\cal O}$ takes a 
value ${\cal O}^{(n)}$ in the $n$th MC step. 

(ii) One may also introduce the rms pair index
\begin{equation}\label{dij}
d_{ij}(\tau) = 
    \sqrt{\{ r_{ij}^2 \} - \{ r_{ij} \}^2} / 
         \{ r_{ij}\}
\end{equation}
to study the MC time evolution. The interest in this quantity 
stems from the following very general observation: 
Consider a configuration space for an $N$-body system which has the
property, that several regions exist where the free energy takes
(local) minima. After a transient time interval 
(``warm up'' or partial equilibration period)  
the MC dynamics starts to explore one of these
minima. There will be a typical time scale (corresponding to an activation energy) 
involved, after which the entire $N$-body system migrates to a 
second, competing minimum  where the procedure repeats itself. 
Thus, the MC dynamics allows to study aspects of the 
energy landscape associated with the configurational space, like
activation barriers. What has been described here for a
general $N$-body system remains equally valid for the 
pair of two particles, Eq. (\ref{dij}), 
embedded in an environment consisting of $N{-}2$ other particles.

\begin{figure}[t]
    \includegraphics[width=0.476\textwidth,clip=true]{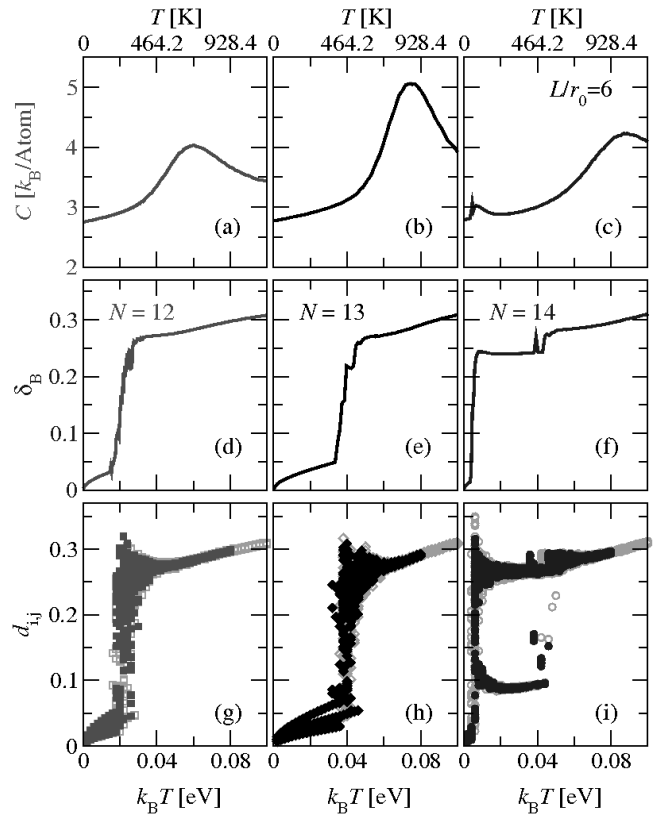}
\caption{\label{CRofT12_14}\sl
  Characteristic temperatures for clusters of
  size $N = 12$ (a)(d)(g), $N = 13$ (b)(e)(h), and $N = 14$
  (c)(f)(i). Panels (a), (b), and (c) show the specific heat,
  panels (d), (e), and (f) the Berry parameter, 
  while panels (g), (h), and (i) show the individual rms bond length fluctuations
  from Eq.\ (\ref{dij}).  
  Panels (a)-(f) show data for $\tau{=}8\times10^7$. 
  In panels (g)-(i) dark symbols
  correspond to $8\times10^7$ MC steps while gray points are
  obtained with $4\times10^7$ MC steps per atom.   
}
\end{figure}

In the limit $\tau{\to}\infty$ the evolution in MC ``time'' is ergodic, 
so the $N$-atom cluster will explore all the phase space available. 
This implies, that at any temperature $T>0$ cluster atoms are
deconfined: there is a finite time after which the 
$i$-atom has migrated from its initial position into any 
other given cluster site. Therefore, 
$d_{ij}$ takes the same value, $d_{\rm B}$
for any given pair of atoms 
and is a unique function of the
temperature and the particle number: $d_{\rm B}(T,N)$. 
Since all diffusion processes terminate at the cluster size
independent of the diffusion constant (i.~e. temperature),
$d_{\rm B}(T,N)$ has only a weak $T$-dependence about a mean value, 
that incorporates crude information about the overall cluster geometry
(spherical vs quasi-onedimensional), but nothing else.
It is implied that the limit 
$d_{\rm B}$ is usually not 
very sensitive to the melting transition.  

For averages like the one defined in Eq. (\ref{dij}) the ergodicity
theorem strictly holds only at $\tau{\to}\infty$; 
at any finite
$\tau$ the value for $d_{ij}(\tau)$ can vary between the different
pairs of atoms $i,j$. If it so happens, that $n$ different classes of
pairs exist, where each class just samples its own local minimum 
in phase space within $\tau$, then 
$d_{ij}(T)$ can develop $n$ branches. 

Quite generally, in situations where different pairs sample different
sectors of phase space, 
the convergence with $\tau$ 
can be increased 
by averaging over all the different pairs.  
In this spirit, we define one more average
\begin{equation}\label{Berry}
\delta_{\rm B}(\tau) = \frac{1}{N(N-1)}\sum_{i,j\neq i} d_{ij}(\tau).
\end{equation}
which we will refer to as the ``Berry parameter'' \cite{ZKBB02}.  
It has a more rapid convergence behavior, $\delta_{\rm
  B}(\tau){\to}d_{\rm B}$, and therefore is easier to
investigate in numerical simulations than the rms pair index.

\subsection*{General properties of $\delta_{\rm B}$ and the rms pair index
  $d_{ij}$}
In order to illustrate the general properties of Berry
parameter and pair index, we now consider as an 
example Al$_{12}$, Al$_{13}$ and Al$_{14}$ (Fig.~\ref{CRofT12_14}).  
We display the Berry
parameter Eq.\ (\ref{Berry}) in Figs.\ \ref{CRofT12_14}(d), (e), (f)
for clusters with $N = 12, 13, 14$. 

At smallest temperatures only thermal vibrations of the atoms
around a single site are observed within the MC window of time.
Hence, the Berry parameter grows $\propto \sqrt T$
reflecting the virial theorem applied to the harmonic oscillator.
This asymptotic low temperature behavior is clearly observed
in the traces Fig.~\ref{CRofT12_14}.

Also easily understood is the behavior
at temperatures higher than $T_\delta$,
at which the Berry parameter exhibits a very sharp jump. 
An estimate of $\delta_\text{B}$ in this regime may be obtained,
by taking the ground state geometry 
and calculating the average squared displacements by assuming ergodicity,
i.e. that in the MC time evolution each occupation
of allowed sites occurs with the same probability. 
For example, in the case of Al$_\text{13}$ one thus finds 
$\delta_\text{B}\approx 0.25$ which agrees reasonably well with
the data, Fig.~\ref{CRofT12_14}.

The sharp increase of $\delta_\text{B}$
at the intermediate temperature, $T_{\delta}$, 
signalizes the
onset of cyclic, correlated exchanges of (surface) atoms between their various 
positions. (The weak irregularities, which are still visible, 
resemble residual statistical noise.) 
In the spirit of our earlier discussion, 
we do not expect that $T_{\delta}$ is independent of our 
observation time $\tau$. \cite{frantz95}
In fact, the precise 
meaning of $T_{\delta}$ is the following: 
at $T_{\delta}$ the observation time $\tau$ has been long 
enough, so that at $T>T_{\delta}$ 
processes can be observed where atoms trade 
places with one another even though the probability $p$ for this 
to happen may be exponentially suppressed with a factor
$\exp(-\Delta/k_{\rm B}T)$. $\Delta$ denotes
the corresponding activation energy which will
in general exhibit a weak (i.e. non-singular)
temperature dependency. 

Our argument shows that $T_{\delta}$ itself 
can not immediately be identified with  
any intrinsic energy scale of the free cluster, like a 
surface melting temperature. The specific heat peaks 
only at a much higher temperature, 
$T_{\rm C}\gg T_{\delta}$, which indicates the  
volume melting temperature of the cluster, 
see Figs.\ \ref{CRofT12_14}(a), (b), (c). 

It is possible to obtain an estimate of the activation energy 
$\Delta$ from the way that $T_\delta$ flows with the
observation time $\tau$.
Namely, one has $ \Delta^{-1}\sim d T_{\delta}^{-1}/d
\ln(\tau)$. Unfortunately, in order to
obtain very accurate scaling with $\ln(\tau)$ the calculational effort goes
well beyond what was achievable within this study.

   \begin{figure}[b]
       \includegraphics[width=0.12\textwidth,clip=true]{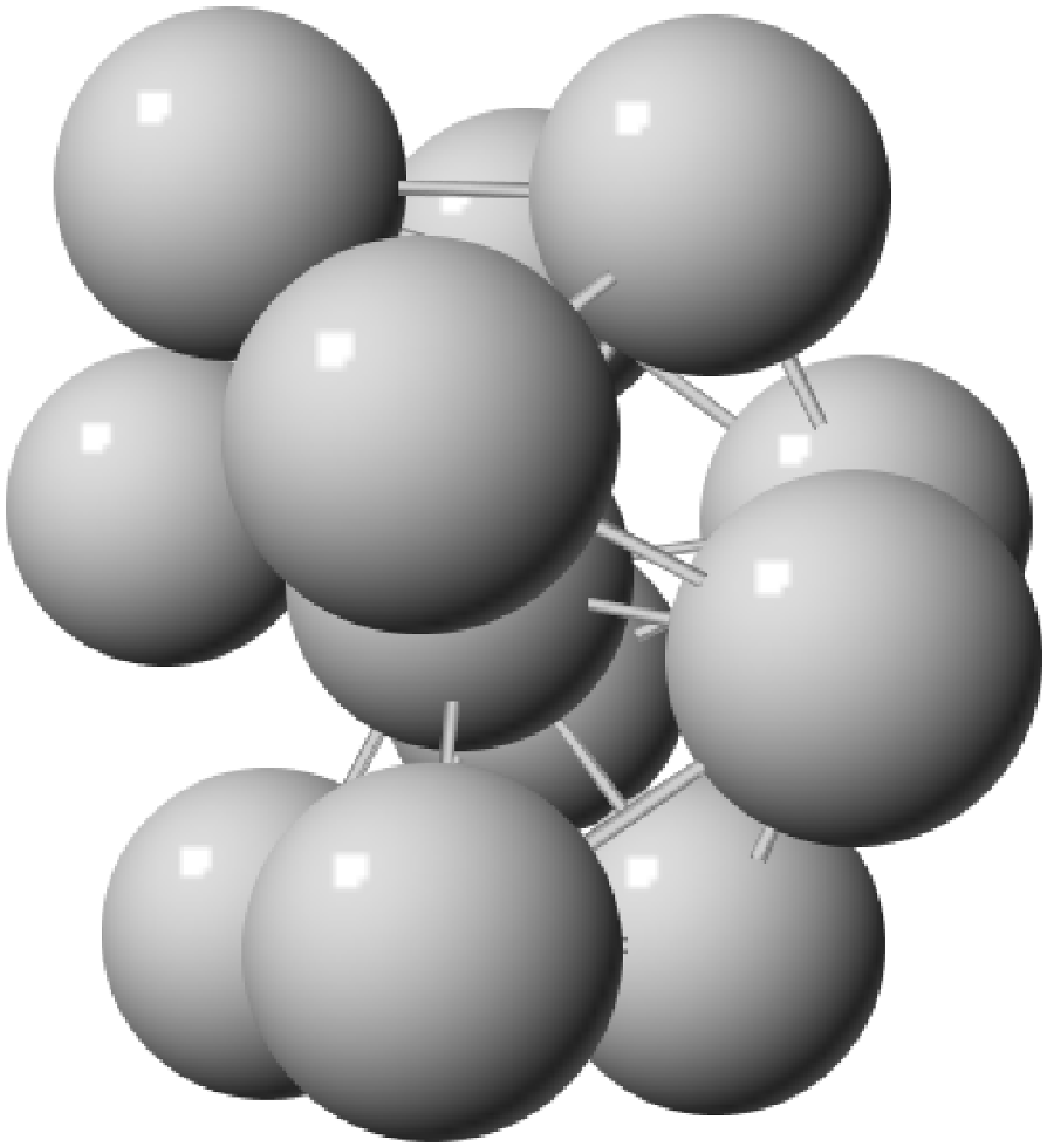}
       \includegraphics[width=0.14\textwidth,clip=true]{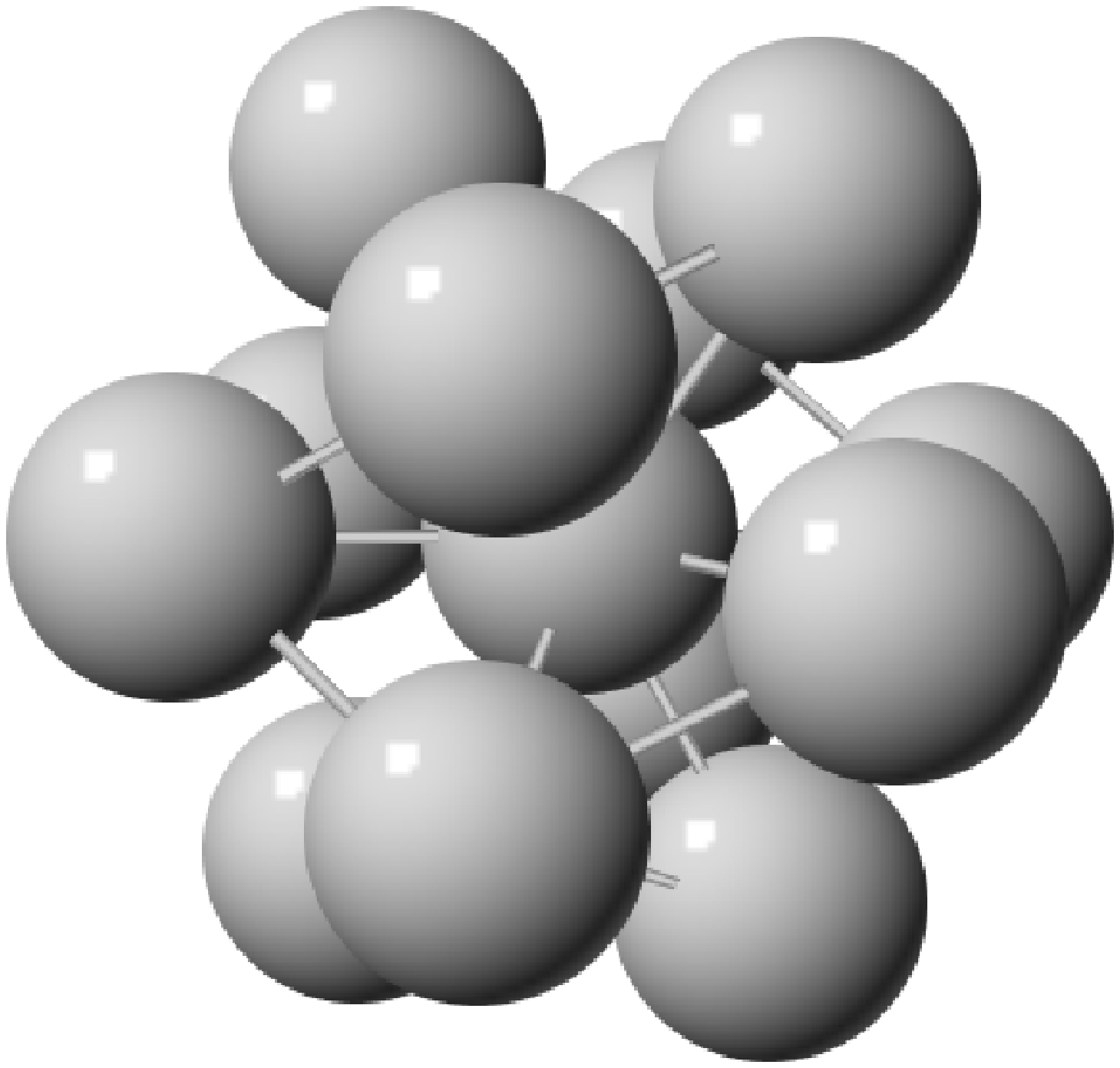}
       \includegraphics[width=0.14\textwidth,clip=true]{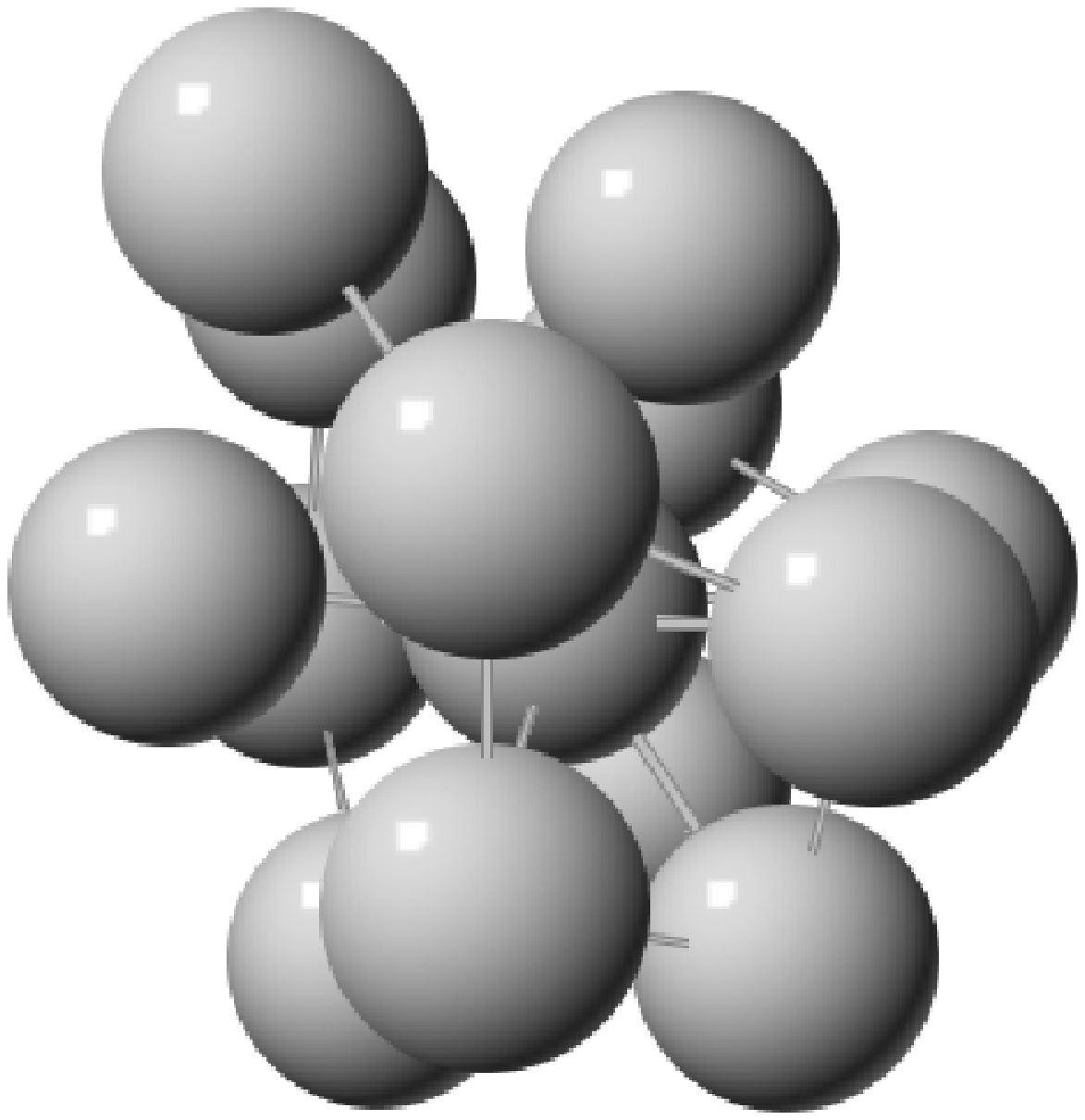}
   \caption{\label{f2}\sl
     Ground state configurations of Al$_{12}$ (left), 
     Al$_{13}$ (center) and Al$_{14}$ (right)
   as obtained with Gupta-potentials. The hole 
   has little impact on the position of the remaining atoms. By
   contrast, the ad-atom creates a docking site with a higher
   coordination number, which is incompatible with the three-fold 
   symmetries of the docking sites of the unperturbed Al$_{13}$. 
   Hence, Al$_{14}$ experiences significant strain.}
   \end{figure}

Further information is carried by the pair index $d_{ij}$, which 
is displayed in Figs.\ \ref{CRofT12_14}(g), (h), (i)
at two values of $\tau$. At $N{=}13$, a three branch structure is
readily identified at $T<T_{\delta}$. The branches reflect the fact, that three 
kinds of atom pairs exist (center atom/shell atom, shell neighbors, shell
next nearest neighbors), that have different distance fluctuations, 
see Fig.~\ref{f2}. 
Two branches are found with $N{=}14$ even at $T>T_{\delta}$. 
The upper one stems from the on shell pairings while the 
lower branch represents the mixed pairs, 
center atom/shell atom. Consistent with this picture, the
lower branch contains 13 bonds, which are 
the 13 bonds between the center atom and the 13
statistically equivalent surface atoms. This
latter branch exists only in the intermediate temperature interval 
$T_{\delta} < T < T_{\rm D}$. 
The role of the temperature $T_{\rm D}$ is similar to $T_{\delta}$, 
except that the associated activation energy, 
$\Delta_{io}$, now is related to an exchange of inner atoms with the outer
shell. At $T_{\rm D}$ the Berry parameter, $\delta_{\rm B}$, 
exhibits a second sharp increase. The discussion of the pair index
allows us to attribute this increase as being due to the  
center atom now being deconfined within the MC window of time.    

   \begin{figure}
       \includegraphics[width=0.476\textwidth,clip=true]{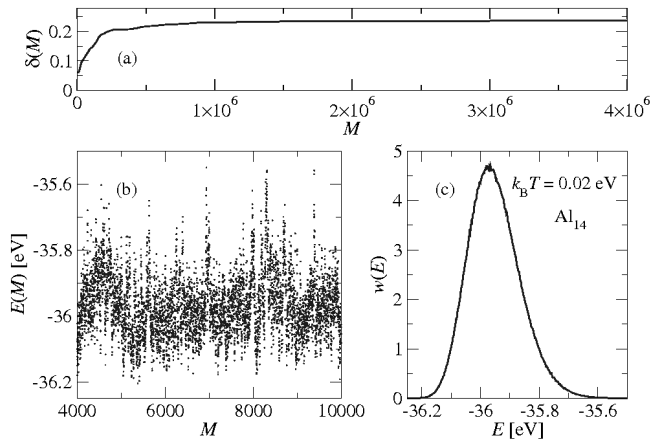}
   \caption{\label{Statistics14}\sl
     Potential energy statistics for Al$_{14}$ at $k_{\rm B}T=0.02$ eV
     in the surface molten phase. (a) Convergence of the Berry
     parameter as a function of MC steps. (b) Sample of the potential
     energies of accepted configurations during the same MC run as in
     (a). (c) Histogram of the potential energy distribution function
     $w(E)$ ($2.4 \times 10^{6}$ energy values, 480 bins). 
   }
   \end{figure}

\begin{table}[t]
\caption{\label{t1} Nomenclature of relevant temperature scales of 
MC simulation with observation time $\tau$ 
and the corresponding activation energies.}
\begin{ruledtabular}
\begin{tabular}{c|l|c}
$T_{\rm \delta}$ & onset of intralayer diffusion observed within $\tau$  & $\Delta$ \\ \hline
$T_{\rm D}$ & onset of interlayer diffusion observed within $\tau$   & $\Delta_{io}$ \\ \hline
$T_{\rm C}$ & maximum of specific heat &
\end{tabular}
\end{ruledtabular}
\end{table}

\section{Self diffusion}\label{sIII}

The influence of closed shells on the cohesive energies of metal
clusters \cite{Wern05a,DW98,AE99,TJW00,WDM+00} in the gas phase 
and their melting points\cite{BB01,WDS+01} has been a focus 
of research for quite some time. 
In this section we use the Berry parameter to investigate
the onset of self diffusion of atoms within such systems.

\subsection{Al$_{14}$}
The low temperature jump in the Berry parameter for Al$_{14}$ 
in Fig.~\ref{CRofT12_14} has been
identified previously as induced by the temporary absorption of the
ad-atom on the Al$_{13}$ icosahedral core structure into the surface, 
see Fig.~\ref{f2}. 
\cite{Wern05a}
Here we witness the effect, that has been described in general terms
already in the introduction. The ad-atom destabilizes the high symmetry
surface of the Al$_{13}$ cluster and therefore the activation barrier
for self diffusion on the surface is reduced, by a factor of roughly  
$T_{\delta}({\rm Al}_{14})/T_{\delta}({\rm Al}_{13}){\sim}5$ according
to our calculation.


In order to obtain a more detailed understanding of
the thermodynamics of the system, in Fig.~\ref{CRofT12_14}(i) the
temperature dependence of the rms pair index is shown. 
In particular, we observe 
that $T_{\rm D}({\rm Al}_{14}){\apprge}T_{\delta}({\rm Al}_{13})$. 
This suggests, that the core atom of Al$_{14}$ must overcome 
a slightly increased barrier (as compared to Al$_{13}$)
to enter the (strained) outer shell.  

For illustration, 
the potential energy statistics for Al$_{14}$ 
are shown in Fig.~\ref{Statistics14} at an 
intermediate temperature ($k_{\rm B}T=0.02$ eV), 
where the pair index exhibits two branches (Fig.~\ref{CRofT12_14}). 
Panel (a) shows the convergence behavior of the Berry
parameter as a function of $\tau$. Panel (b) displays potential
energies of accepted configurations during 
the sample MC run. Panel (c) shows the potential energy distribution
function, $w(E)$. Consistent with our interpretation of the branching
behavior and with earlier results for Ni$_{14}$ the distribution
shows no sign of phase coexistence. \cite{KB93,KB94,PPB01} 

The results presented here for Al$_{14}$ are analogous to those for
Pb$_{14}$ with very similar characteristic temperatures
\cite{LLKN01}. 
They differ from Ni$_{14}$ \cite{LML+00} and Cu$_{14}$ \cite{lopez95}
in so far, as for Al$_{14}$ no ad-atom hopping is observed. 

\begin{figure}
       \includegraphics[width=0.476\textwidth,clip=true]{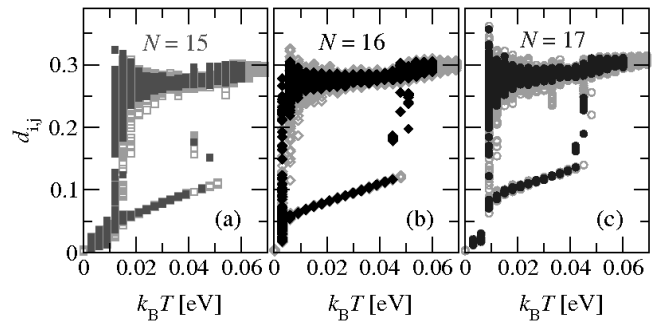}
   \caption{\label{CRofT15_17}\sl
     Temperature dependence of the individual bond fluctuations Eq.\
     (\protect\ref{dij}) for clusters of size $N = 15$ (a), $N = 16$
     (b), and $N = 17$ (c). The ``solid'' bonds with $d_{ij} \le 0.1$
     correspond to those between the central atom and the surface
     atoms. In the surface molten phase all surface atoms are
     equivalent. Dark symbols correspond to $8\times10^7$ MC steps
     while gray open symbols are obtained with $2\times10^7$ MC steps per
     atom. 
   }
   \end{figure}

\begin{figure*}
       \includegraphics[width=0.8\textwidth,clip=true]{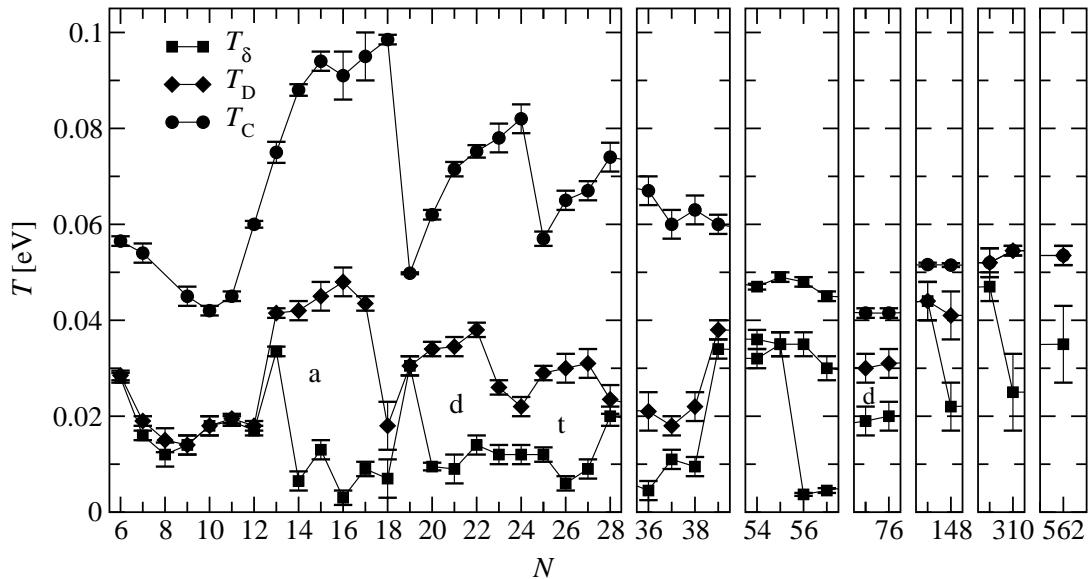}
   \caption{\label{PhaseDiagram}\sl
     Characteristic temperatures of Al$_{N}$ associated with self
     diffusion over the cluster size $N$; 
     for the definitions of $T_{\delta},T_{\rm D}, T_{\rm C}$ 
     see Tab.~\ref{t1}. 
     "a" labels a ``confined'' central atom, 
     "d" a ``confined'' central
     dimer, and "t" a ``confined'' central trimer. 
(The onset of the evaporation transition in the
investigated  systems is found at larger temperatures 
$T_{\rm evap} > 0.12$ eV \cite{Wern05a}.)
   }
   \end{figure*}

The clusters with $15\leq N\leq 18$ show a behavior similar to Al$_{14}$. 
Apparently, the presence of several additional ad-atoms has
qualitatively a similar (destabilizing) effect on the surface 
as a single ad-atom. Panels (a), (b), and (c) of Fig.~\ref{CRofT15_17} show the individual 
bond length fluctuations Eq.\ (\ref{dij}) for Al$_{15-17}$, 
respectively. 
For Al$_{18}$ (not shown here, c.f.\ Figs.\ \ref{PhaseDiagram}) 
similar results are obtained.

\subsection{Al$_{56}$ and Al$_{57}$}
In order to see whether the lowering of activation
barriers for surface diffusion may indeed be 
a typical phenomenon for closed
shell configuration with one excess atom, 
we now investigate the case Al$_{56}$. 

\begin{figure}
       \includegraphics[width=0.476\textwidth,clip=true]{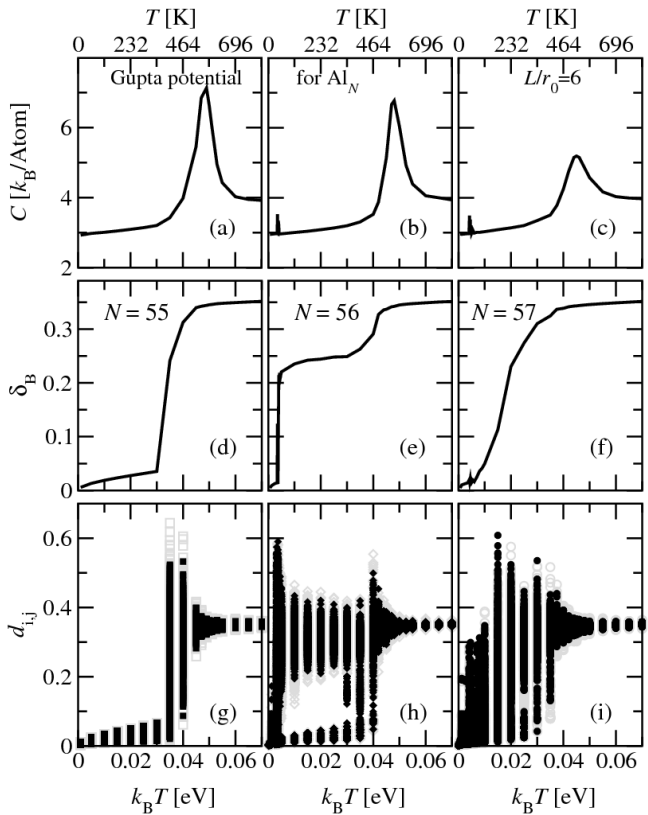}
   \caption{\label{CRofT54_56}\sl
     Temperature dependence of the melting transition for clusters of
     size $N = 55$ (a)(d)(g), $N = 56$ (b)(e)(h), and $N = 57$
     (c)(f)(i). Panels (a), (b), and (c) show the specific heat,
     panels (d), (e), and (f) the Berry parameter, while panels (g),
     (h), and (i) show the individual rms bond length fluctuations
     from Eq.\ (\ref{dij}). Panels (a)-(d) show data for $8\times10^7$
     MC steps per atom; in panels (g)-(i) dark symbols
     correspond to $8\times10^7$ MC steps while gray open symbols are
     obtained with $4\times10^7$ MC steps per atom. 
   }
   \end{figure}

Figure
\ref{CRofT54_56} shows the specific heat, the Berry parameter and the
pair index for $N=55$, $56$, and $57$, respectively.  
The dramatic suppression of $T_{\delta}$ in Al$_{56}$ seen in
$\delta_{\rm B}$ as compared to the closed shell
case Al$_{55}$ was already observed before\cite{Wern05a}. 

   \begin{figure}[b]
       \includegraphics[width=0.476\textwidth,clip=true]{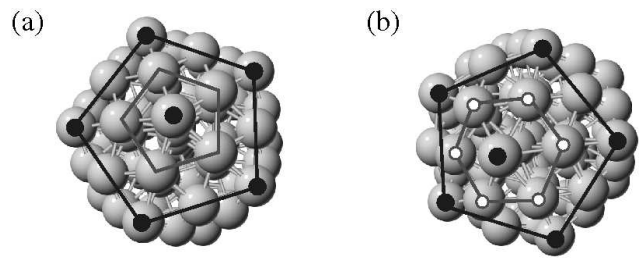}
   \caption{\label{GS55_56}\sl
     Ground state configurations of Al$_{55}$ (a) and Al$_{56}$
   (b)as obtained with Gupta-potentials. 
   Large full circles indicate the corner atoms, which remain
   ``solid'' in the partially surface molten state of
   Al$_{56}$. Small open circles (rosette structure \cite{apra04})
   show the broken fivefold symmetry
   caused be the 56$^{\rm th}$ atom absorbed into the surface of the
   icosahedron.
   }
   \end{figure}

To obtain more detailed information, we consider the pair index
Fig.~\ref{CRofT54_56}(i). The lower branch at $T_\delta < T < T_{\rm D}$ 
represents 331 pairs. They correspond to all pairings within the 
set of atoms that consists of the Al$_{13}$ core together with  
the 12 outer corner atoms of the (distorted) icosahedron -- 
subset of confined atoms. The remaining 31 surface
atoms are deconfined. The upper branch represents the cross-pairings
between the  two sets of atoms as well as the pairings between
deconfined atoms. The data suggests, that edge atoms of surface facets 
are more mobile than the corner atoms. 

In order to understand the
reason for this enhanced mobility, 
we display in Figure \ref{GS55_56} the (Gupta-potential based) 
ground state structures 
of Al$_{56}$ and Al$_{55}$. The large full circles
indicate the 12 corner atoms, which move very little at
$T_\delta<T<T_{\rm D}$. 
It is seen, that the ad-atom integrates by promoting the environment of one
of the corner atoms from its original 5 fold symmetry to an
(approximate) local 6 fold symmetry. This implies a
sizable bond length mismatch of about $1/5{\sim}20$\%, that creates a
deformation field. Now, if an atom out of the six 
ring pushes away another atom to enter a new neighboring facet, 
then the entire deformation field will follow. Actually, 
what really diffuses over the surface 
is the six-fold rosette structure (Fig.~\ref{GS55_56}). 
The energy barrier to be overcome in this self 
diffusion process is relatively low. This is 
because it is not necessary to first create a hole 
in the crystal lattice of the target facet for 
the rosette to move there. 

According to the scenario developed for Al$_{56}$,
two rosettes should decorate the surface of Al$_{57}$.
A naive expectation is, that these rosettes repel each another, because
it is more difficult for a second rosette to enter an area that is under
strain already from the presence of a first one. Since the two
rosettes cover a large surface fraction of Al$_{57}$, diffusion
barriers should be increased again as compared to the "free" case
Al$_{56}$. Indeed, as can be seen in Fig.~\ref{CRofT54_56},
the sharp increase observed for Al$_{56}$ at very low temperatures
in the Berry parameter almost disappears for Al$_{57}$.
It gives way to a very broad shoulder which is indicative of a large
number of energy scales that is associated with a strongly disturbed
("disordered") outer cluster shell.  

{\it Experiments:} In a recent experiment, a broad peak in the specific heat, $C(T)$, of
Al$_{56}^{+}$  and Al$_{57}^{+}$
has been measured and analyzed.\cite{BNCJ05,neal07}
The authors were concluding, 
that Al$_{56}^{+}$ undergoes a separate transition even before the entire cluster
starts to melt. The physical nature of the first transition could not be
clearly resolved. It would be tempting to propose that our research 
supports the supposition of the authors that
premelting of the surface is a possible candidate. 
Namely, since low diffusion barriers are usually also indicative of low
melting temperatures, our scenario would
suggest that premelting of the 
surface may occur well before melting of the bulk in
Al$_{56}^{+}$.
However, such a direct application of our ideas to
experiments\cite{BNCJ05,neal07} is not without difficulty and
probably not indicated: there is a
trend in the theoretical data, Fig.~\ref{CRofT54_56},
that the latent heat 
(integral under the peak in the specific heat)
decreases from $N{=}55$ to $N{=}57$. This is plausible, because
it takes less energy per atom to melt a structure under strain.
By contrast, the experimental trend is reverse and the latent heat
increases for the series $N{=}55,56,57$. 
The likely reason for this discrepancy is that
Gupta-potentials cannot give a 
sufficiently realistic description of the
thermodynamics of aluminum clusters, Al$_{N}^{+}$,
($N{=}55,56,57$) near the melting transition. 
That indeed difficulties exist even with the uncharged species,
Al$_{N}$, is signalized by explicit density functional theory
calculations, which  show that 
the $T{=}0$ ground state conformation of Al$_{\rm 55}$ 
is not icosahedral \cite{AE99,aguado06} contradicting what is 
found with Gupta potentials. This casts a  doubt on the
applicability of Gupta-potentials to the specific metal aluminum
for simulations of low temperature behavior. However, we would like
to emphasize that our general findings remain valid for other
materials with an isocahedral ground state as well as for those
situations, where an approximate icosahedral symmetry is restored at
slightly higher temperatures.

\section{Characteristic temperatures for Al$_{N}$-clusters}\label{sIV}

The delicate interplay between the special geometry of Al$_{55}$ and the 
low activation barrier for ad-atom diffusion in Al$_{56}$ and
Al$_{57}$ does not prompt the
expectation, that this specific mechanism is ubiquitous in all its details. 
Nevertheless, there is a lesson to be learned about the more general case.
One expects a lowering of the activation energy for surface diffusion
of surface atoms of clusters with structures that derive from  
high symmetry parent states either by (i) punching in vacancies or  
(ii) by inserting ad-atoms into its outer shell.  
In such systems, there is a possibility for atomic motions, which are not 
just ring exchanges and which also do not require to break chemical bonds
to first create a vacancy. 
Since low diffusion barriers are also indicative of 
low melting temperatures, here is a mechanism by which surface melting 
may become a process that should be distinguished, in principle,  
from the melting of the bulk. 

This analysis suggests, that the ratio
$T_{\delta}/T_{\rm D}$ tends to be large for structures with 
closed shells or subshells and much smaller otherwise. 
We have tested this idea by calculating $T_{\rm C}$,
the temperature at which the specific heat $C(T)$ takes 
its maximum, $T_\delta$ and $T_{\rm D}$ (at fixed $\tau$) for 
a variety of different cluster sizes. 
Figure \ref{PhaseDiagram} comprises our results 
which, we believe, support our general picture:  
the ratio of $T_\delta/T_{\rm D}$
takes peak values at closed shell structures and much lower ones
almost everywhere
else (except for the smallest cluster sizes, where our previous analysis
does not apply).

We mention that the Al$_N$ clusters with $14 \le N \le 18$ and $N=24$ have maxima of
the specific heat at temperatures larger than the Al bulk melting temperature of 
$T_{\rm bulk} = 933\ {\rm K} = 0.0804\ {\rm eV}/k_{\rm B}$. This observation is in 
line with the empirical investigation of the melting of small 
Sn\cite{SJ00} and 
Ga\cite{BBS+03} 
clusters, which have revealed a possible
stability of the solid phase of the particles beyond the melting
temperature of the bulk material. In these cases the high
cluster melting temperature was interpreted as a
consequence of the  rigidity of the specific
ground state structures of the
clusters. This interpretation found support from
microcanonical molecular dynamics (MD) 
calculations for C, Si, Ge, and Sn 
clusters\cite{LWH00} 
as well as for isokinetic MD
investigations of 
Sn$_{10}$\cite{JKB02,BNCJ05} and 
Ga$_{13}$, Ga$_{17}$\cite{CJKB04} 
particles.

We briefly touch upon the limit of large clusters, $N\gg 100$. 
There, we observe that $T_{\rm D}{\to} T_{\rm C}$, 
while $T_\delta$ does not appear to follow this trend. 
To understand this behavior, recall that $T_{\rm D}$ is the temperature at
which our MC time has become long enough, so we can observe an 
exchange of particles between the outer shell and its inner neighbor. 
Then, $T_{\rm D}{\approx}T_{\rm C}$ implies that intershell exchange cannot 
be observed -- even with our very long observation times --
unless we actually heat up the entire cluster to melt. 
This behavior is consistent with
what one would expect for a macroscopic single crystal grain: 
as long as the crystalline structure of the surface is intact
($T\ll T_{\rm melt}$) one has $\Delta\ll \Delta_{io}$ and
interlayer-diffusion is strongly suppressed. 
However, after melting the surface layer no longer forces the atoms of
its neighboring inner shell into a crystalline structure. Hence, 
intra self diffusion within the second layer becomes (almost) 
as cheap energetically as was diffusion in the first layer, 
before. Therefore, the second layer melts immediately 
after the first one, so at melting 
one has for the ``effective'' activation energies: 
$\overline \Delta\approx \overline \Delta_{io}\sim T_{\rm melt}$. 
In other words: we recover the standard ``continuous melting'' 
scenario. 

\section{Conclusion}\label{sConclusion}

In this paper we have argued, that certain metal clusters with an intermediate size
may exhibit a property that cannot be found in bulk materials: the 
activation energies for surface diffusion, $\Delta$, and interlayer
diffusion, $\Delta_{io}$, are substantially different from one another:
$\Delta_{io}\gg\Delta$. Since these activation energies are closely tied to
melting temperatures, one expects that the outermost surface layer 
can exhibit its own melting transition, which is well separated from the
bulk. The continuous melting of the cluster core should start only 
at much higher temperatures. 

The most dramatic decrease of $\Delta$ has been found with Al$_{56}$. 
In this case, it is the ``frustration'' of atomic bonds that originates from
implanting an ad-atom into a closed shell system, which produces the effect. 
A related mechanism leads to a decrease in the surface melting
temperature also for  Al$_{14}$ and is expected to be active in
clusters, where $N$ is slightly above some magical (closed atom shell)
value $N^*$. We mention, that a distortion of the outer cluster shell is
also present in clusters with $N$ slightly below $N^*$. However, a
hole is usually accommodated more easily than an ad-atom and therefore
the decrease tends to be asymmetric: it is typically 
stronger for $N>N^*$ as compared to $N<N^*$.    

Monatomic, macroscopic and planar metal surfaces, that face the vacuum, 
do not easily allow for the frustration of surface bonds that we have
observed with the metal clusters. An attempt to locally implant
concentration of ad-atoms into macroscopic surfaces beyond a certain
threshold would result in a metastable state that eventually would 
transform into another state without frustration, where the ad-atoms would 
have undergone island formation. 

On the other hand, our research suggests the design of materials with
a surface melting temperature that is strongly diminished and separates 
from the bulk melting temperature by a controllable amount. The idea
is to employ a monatomic core and a biatomic shell structure. The purpose of pressing  
foreign atoms into the outermost layer of the host material's crystal is to
create local strain fields. Since strain reduces the local melting
temperature, heating up such a system could create puddles of molten host
material on top of the solid, bulk core. Clearly, the combination of host and 
implantation materials  should satisfy at least two conditions: 
(i) the implantations should have a high solubility in the host material, 
but (ii) they should not easily diffuse away from the surface into the bulk of 
the crystal, either. 
Whether indeed a combination of suitable materials can be found, this
we have to leave for future research. 

\subsection{Acknowledgments}
Instructive discussions with M.\ Blom and P.\ Schmitteckert
are gratefully acknowledged. The work was
supported by the Center for Functional Nano\-struc\-tures of the
Deutsche Forschungsgemeinschaft within project C4.7.



\begin{thebibliography}{10}

\bibitem{hakkinen03}
H. H\"akkinen, S. Abbet, A. Sanchez, U. Heiz, and U. Landman, Angew. Chem. Int.
  Ed. {\bf 42},  1297  (2003).

\bibitem{catalysis03}
S. Rojas, S. Eriksson and M. Boutonnet,
{\it Microemulsion: an Alternative Route to Preparing Supported
Catalysts},  in {\it Catalysis}, Vol. 17, Special Periodical Reports,
Royal Society of Chemistry (2004); P. L. Gai and E.D. Boyes,
{\it Electron Microscopy in Heterogeneous Catalysis}, Chapt. 5,
Series in Microscopy in Materials Science, Institute of Physics
(2003). 
  
\bibitem{gross01}
 D. Gross, {\it Microcanonical Thermodynamics: Phase Transitions in
 "Small" Systems}, World Scientific Publishing Co. Pts. Ltd, Singapore,
(2001).

\bibitem{AB86}
F.~G. Amar and R.~S. Berry, J. Chem. Phys. {\bf 85},  5943  (1986).

\bibitem{JBB86}
J. Jellinek, T.~L. Beck, and R.~S. Berry, J. Chem. Phys. {\bf 84},  2783
  (1986).

\bibitem{KJ97}
E.~B. Krissinel and J. Jellinek, Int. J. Quant. Chem. {\bf 62},  185  (1997).

\bibitem{WDM+00}
D.~J. Wales, J.~P.~K. Doye, M.~A. Miller, P.~N. Mortenson, and T.~R. Walsh,
  Adv. Chem. Phys. {\bf 115},  1  (2000).

\bibitem{BF05}
F. Baletto and R. Ferrando, Rev. Mod. Phys. {\bf 77},  371  (2005).

\bibitem{GS70}
R.~M. Goodman and G.~A. Somorjai, J. Chem. Phys. {\bf 52},  6325  (1970).

\bibitem{cahn86}
R.~W. Cahn, Nature {\bf 323},  668  (1986).

\bibitem{Dash02}
J.~G. Dash, Contemporary Physics {\bf 43},  427  (2002).

\bibitem{PGFV87}
B. Pluis, A.~D. van~der Gon, J. Frenken, and J. van~der Veen, Phys. Rev. Lett.
  {\bf 59},  2678  (1987).

\bibitem{FV85}
J.~W.~M. Frenken and J.~F. van~der Veen, Phys. Rev. Lett. {\bf 54},  134
  (1985).

\bibitem{OLW94}
R. Ohnesorge, H. L\"owen, and H. Wagner, Phys. Rev. E {\bf 50},  4801  (1994).

\bibitem{TET95}
F.~D. DiTolla, F. Ercolessi, and E. Tosatti, Phys. Rev. Lett. {\bf 74},  3201
  (1995).

\bibitem{DW98}
J.~P.~K. Doye and D.~J. Wales, New. J. Chem. {\bf 1998},  733  (1998).

\bibitem{AE99}
R. Ahlrichs and S.~D. Elliot, Phys. Chem. Chem. Phys. {\bf 1},  13  (1999).

\bibitem{TJW00}
G.~W. Turner, R.~L. Johnston, and N.~T. Wilson, J. Chem. Phys. {\bf 112},  4773
   (2000).

\bibitem{aguado06}
A. Aguado and J.~M. Lopez, J. Phys. Chem. B {\bf 110}, 1402 (2006). 

\bibitem{noya07}
E.~G. Noya, J.~P.~K. Doye, D.~J. Wales and A. Aguado, 
Eur. Phys. J. D {\bf 43}, 57 (2007). 

\bibitem{KB93}
R.~E. Kunz and R.~S. Berry, Phys. Rev. Lett. {\bf 71},  3987  (1993).

\bibitem{haberland05}
H. Haberland, T. Hippler, J. Donges, O. Kostko, M. Schmidt, and B. von
  Issendorff, Phys. Rev. Lett. {\bf 94},  035701  (2005).

\bibitem{schmidt98}
M. Schmidt, R. Kusche, B.~V. Issendorff, and H. Haberland, Nature {\bf 393},
  238  (1998).

\bibitem{aguado05c}
A. Aguado, J. Phys. Chem. B, {\bf 109} 13043 (2005). 

\bibitem{aguado05a}
A. Aguado and J.~M. Lopez, Phys. Rev. Lett. {\bf 94}, 233401 (2005). 

\bibitem{krishnamurty07}
 S. Krishnamurty, G.~S. Shafai, D.~G. Kanhere, B.~S. de Bas and
 M.~J. Ford, J. Phys. Chem. A {\bf 111}, 10769 (2007). 

\bibitem{BNCJ05}
G.~A. Breaux, C.~M. Neal, B. Cao, and M.~F. Jarrold, Phys. Rev. Lett. {\bf 94},
   173401  (2005).

\bibitem{neal07}
C.~M. Neal, A.~K. Starace, and M.~F. Jarrold, Phys. Rev. B {\bf 76},  054113
  (2007).

\bibitem{Wern05a}
R. Werner, Eur.\ Phys. J.\ B {\bf 43},  47  (2005).

\bibitem{Gupt81}
R.~P. Gupta, Phys. Rev. B {\bf 23},  6265  (1981).

\bibitem{TMB83}
D. Tom\'anek, S. Mukherjee, and K.~H. Bennemann, Phys. Rev. B {\bf 28},  665
  (1983).

\bibitem{ZLT91}
W. Zhong, Y.~S. Li, and D. Tom\'anek, Phys. Rev. B {\bf 44},  13053  (1991).

\bibitem{CR93}
F. Cleri and V. Rosato, Phys. Rev. B {\bf 48},  22  (1993).

\bibitem{AT89}
M.~P. Allen and D.~J. Tildesley, {\it Computer Simulation of Liquids}, {\it
  Oxford Science Publications} (Clarendon Press, Oxford, 1989).

\bibitem{WDS+01}
J. Wang, F. Ding, W. Shen, T. Li, G. Wang, and J. Zhao, Solid State Commun.
  {\bf 119},  13  (2001).

\bibitem{ZKBB02}
Y. Zhou, M. Karplus, K.~D. Ball, and R.~S. Berry, J. Chem. Phys. {\bf 116},
  2323  (2002).

\bibitem{frantz95}
D.~D. Frantz, J. Chem. Phys. {\bf 102},  3747  (1995).

\bibitem{BB01}
P. Blaise and S.~A. Blundell, Phys. Rev. B {\bf 63},  235409  (2001).

\bibitem{KB94}
R.~E. Kunz and R.~S. Berry, Phys. Rev. E {\bf 49},  1895  (1994).

\bibitem{PPB01}
A. Proykova, S. Pisov, and R. Berry, J. Chem. Phys. {\bf 115},  8583  (2001).

\bibitem{LLKN01}
Y.~J. Lee, E.-K. Lee, S. Kim, and R.~M.~Nieminen, Phys. Rev. Lett. {\bf 86},
   999  (2001).

\bibitem{LML+00}
Y.~J. Lee, J.~Y. Maeng, E.-K. Lee, B. Kim, S. Kim, and K.-K. Han, J. Comput.
  Chem. {\bf 21},  380  (2000).

\bibitem{SJ00}
A.~A. Shvartsburg and M.~F. Jarrold, Phys. Rev. Lett. {\bf 85},  2530  (2000).

\bibitem{BBS+03}
G.~A. Breaux, R.~C. Benirschke, T. Sugai, B.~S. Kinnear, and M.~F. Jarrold,
  Phys. Rev. Lett. {\bf 91},  215508  (2003).

\bibitem{LWH00}
Z.-Y. Lu, C.-Z. Wang, and K.-M. Ho, Phys. Rev. B {\bf 61},  2329  (2000).

\bibitem{JKB02} 
K. Joshi, D.~G. Kanhere, and S.~A. Blundell, Phys. Rev. B {\bf 66},  155329
  (2002).

\bibitem{CJKB04}
S. Chacko, K. Joshi, D.~G. Kanhere, Phys. Rev. Lett. {\bf
  92},  135506  (2004).


\bibitem{lopez95} 
M.~J. Lopez, P.~A. Marcos and J.~A. Alonso, J. Chem. Phys. {\bf 104},
1056 (1995). 

\bibitem{apra04}
E. Apra, F. Baletto, R. Ferrando and A. Fortunelli,
Phys. Rev. Lett. {\bf 93}, 065502 (2004). 

\end{thebibliography}


\end{document}